\documentclass[showpacs,amssymb,amsmath,nobibnotes, aps,prl,secnumarabic
,superscriptaddress
,twocolumn%
]{revtex4-1}
\pdfoutput=1

\usepackage{bm}
\usepackage{natbib}
\usepackage{graphicx}
\usepackage{color}

\begin{document}
\textfloatsep 10pt

\title{Three-dimensional fluid motion in Faraday waves: creation of vorticity and generation of two-dimensional turbulence}
\author{N. Francois}
\email{Nicolas.Francois@anu.edu.au}
\affiliation{Research School of
Physics and Engineering, The Australian National
University, Canberra ACT 0200, Australia}

\author{H. Xia}
\affiliation{Research School of
Physics and Engineering, The Australian National
University, Canberra ACT 0200, Australia}
\author{H. Punzmann}
\affiliation{Research School of
Physics and Engineering, The Australian National
University, Canberra ACT 0200, Australia}
\author{S. Ramsden}
\affiliation{National Computational Infrastructure (NCI) Vizlab, The Australian National
University, Canberra ACT 0200, Australia}
\author{M. Shats}
\affiliation{Research School of
Physics and Engineering, The Australian National
University, Canberra ACT 0200, Australia}

\date{\today}

\begin{abstract}


\end{abstract}


\maketitle

Popular summary:\\

Parametrically excited waves are a ubiquitous phenomenon observed in a variety of physical contexts. They span from Faraday waves on the water surface to spin waves in magnetics, electrostatic waves in plasma and second sound waves in liquid helium. Parametrically excited Faraday waves on the surface of vertically vibrated liquids quickly become nonlinear. In dissipative liquids, or in granular media, these nonlinear waves form regular lattices of oscillating solitons (oscillons), resembling in some aspects 2D crystals. If the vertical acceleration is increased, the oscillons do not solely grow in amplitude, their horizontal mobility is also greatly enhanced, and ultimately the lattice melts and becomes disordered. Until recently, the physics of these self-organized waves and their transition to disorder have been studied almost exclusively based on the analysis of the wave motion rather than the motion of their constitutive components, whether they are solid grains or fluid particles.
 
  It has recently been discovered that the fluid motion on a liquid surface perturbed by Faraday waves reproduces in detail the statistics of two-dimensional turbulence. This unexpected discovery shifts the current paradigm of order to disorder transition in this system: instead of considering complex wave fields, or wave turbulence, it is conceivable that the 2D Navier-Stokes turbulence, generated by Faraday waves, feedbacks on the wave crystal and disorders it in a statistically predictable fashion. To date, the very mechanism behind the turbulence generation in such waves remains unknown. A better understanding of this phenomenon is important for a wide spectrum of physics applications involving parametric waves.

In this paper, we visualize 3D trajectories of floating tracers and reveal that the fluid particles motion injects 2D vortices into the horizontal flow.
This is an unexpected and new paradigm for vorticity creation in a 2D flow. The horizontal energy is then spread over the broad range of scales by the turbulent inverse energy cascade. Two-dimensional turbulence destroys the geometrical order of the underlying lattice. The crystal order, however, can be restored by increasing viscous dissipation in the fluid which hinders vorticity creation and thus the development of turbulence.\\


Abstract:\\

  We study the generation of 2D turbulence in Faraday waves by investigating the creation of spatially periodic vortices in this system. Measurements which couple a diffusing light imaging technique and particle tracking algorithms allow the simultaneous observation of the three-dimensional fluid motion and of the temporal changes in the wave field topography.
  
	Quasi-standing waves are found to coexist with a spatially extended fluid transport. More specifically, the destruction of regular patterns of oscillons coincides with the emergence of a complex fluid motion whose statistics are similar to that of two-dimensional turbulence.
 We reveal that a lattice of oscillons generates vorticity at the oscillon scale in the horizontal flow. The interaction of these vortices explain how 2D turbulence is fueled by almost standing waves.
 Remarkably, the curvature of Lagrangian trajectories reveals a "footprint" of the forcing scale vortices in fully developed turbulence. 2D Navier-Stokes turbulence should be considered a source of disorder in Faraday waves. These findings also provide a new paradigm for vorticity creation in 2D flows.

\normalfont

\newpage

\section{I. INTRODUCTION}

Parametrically excited waves are observed in many physical systems: Faraday waves on the surface of liquids \cite{Faraday1831}, spin waves in ferrites \cite{Krutsenko1978}, second sound waves in liquid helium \cite{Rinberg1997,Rinberg2001}, and electrostatic waves in plasma \cite{Nishikawa1968}.
On a liquid surface, parametric waves can also be excited using electric fields \cite{Rodishevskii1988}, ultrasonic excitation \cite{Holt1996}, wave paddles \cite{Falcon2009}, plungers \cite{Brazhnikov2002,XiaShats2012}, etc.

Parametric surface waves can self-organize into various motifs and they have been the focus of the pattern formation physics for many years \cite{Cross1993,Douady1990,Gluckman1995}. Stable patterns, crystals, or even quasi-crystals are produced on the surface of dissipative liquids \cite{Kudrolli1996,Kityk2005,Fauve1994} or granular media \cite{Umbanhowar1996,Goldman2003}. At higher wave amplitudes such patterns break down into disordered lattices. Different pathways leading to disorder have been described such as: defect-mediated turbulence, oscillatory transition phase, or the lattice melting scenario  \cite{Gollub1989,Ciliberto1984,Shani2010,Goldman2003,Fineberg2000}.

Until recently the physics of these parametrically excited waves, and their transition to disorder, have been studied almost exclusively based on the analysis of the wave motion rather than the motion of their constitutive solid grains or fluid particles.
Nevertheless recent studies of the motion of floaters on the surface of Faraday waves (FW) revealed an unexpected physics picture: the particle motion, in these three-dimensional waves, reproduces in detail the statistics of two-dimensional turbulence. Both the inverse energy cascade and the spectral condensation have been found, suggesting that Navier-Stokes two-dimensional turbulence could be a source of disorder at work in low dissipation systems \cite{Kameke2011,Francois2013}.

Although FW are quasi-standing waves, early studies already pointed out that they can generate a random fluid particle motion \cite{Gollub1990,Mesquita1992,Hansen1997}. It was suggested that this Lagrangian motion can be modelled by an extension of the Stokes drift phenomenology for random wave field \cite{Mesquita1992,Stokes1847,Hasselmann1982}. However, such a phenomenology shares almost no connection with the recent observation of FW driven two-dimensional turbulence \cite{Kameke2011,Francois2013,XiaNC2013}. Indeed this modelling approach describes poorly the presence of a spectrally localized horizontal forcing mechanism for 2D turbulence. Moreover, it contains no ingredient to capture the existence of an inertial range.
Thus, what remains unknown is how and why waves produce 2D turbulence. In particular, how the energy is injected into horizontal flow from vertical oscillations and why it is injected in a narrow range of scales leaving it to turbulence to spread this energy over a broad inertial interval.

	In this work, we address some of these shortcomings and describe previously unexplored Lagrangian features of particle trajectories in disordered and ordered Faraday waves. We use a combination of a high speed camera, a diffusing light imaging technique and tracking algorithms to reconstruct the three-dimensional (3D) trajectories of floating tracers as well as the topography of the wave field. This approach gives new insights on how an essentially vertical energy injection is converted into spatially extended horizontal fluid motion.
	We confirm a clear distinction between the wave and the fluid motion in disordered FW \cite{Francois2013}. The waves motion is equivalent to that of an assembly of oscillons confined to the sites of a weakly irregular lattice. In this sense, even disordered FW remain close to standing waves. Conversely, particles motion is highly erratic and spatially extended. We show that disordered quasi-standing waves generate vorticity in the horizontal fluid motion. Those vortices are determined by a unique characteristic length scale, and are the fuel of the 2D turbulent flow. As demonstrated in this paper, a substantial increase in viscosity results in the recovery of a perfect standing wave crystal while the particle motion becomes restricted to closed orbits.
	The emerging physical model of Faraday ripples consists of a spatially periodic 2D array of oscillating solitons \cite{Umbanhowar1996,Shats2012} which contain vertical oscillatory energy but also inject horizontal vortices into the flow at the scale of the oscillon size. This horizontal energy is then spread over the broad range of scales by the inverse energy cascade and at the same time 2D turbulence disorders the underlying lattice.

\section{II. RESULTS}

\begin{figure*}[!t]
\includegraphics[width=18 cm]{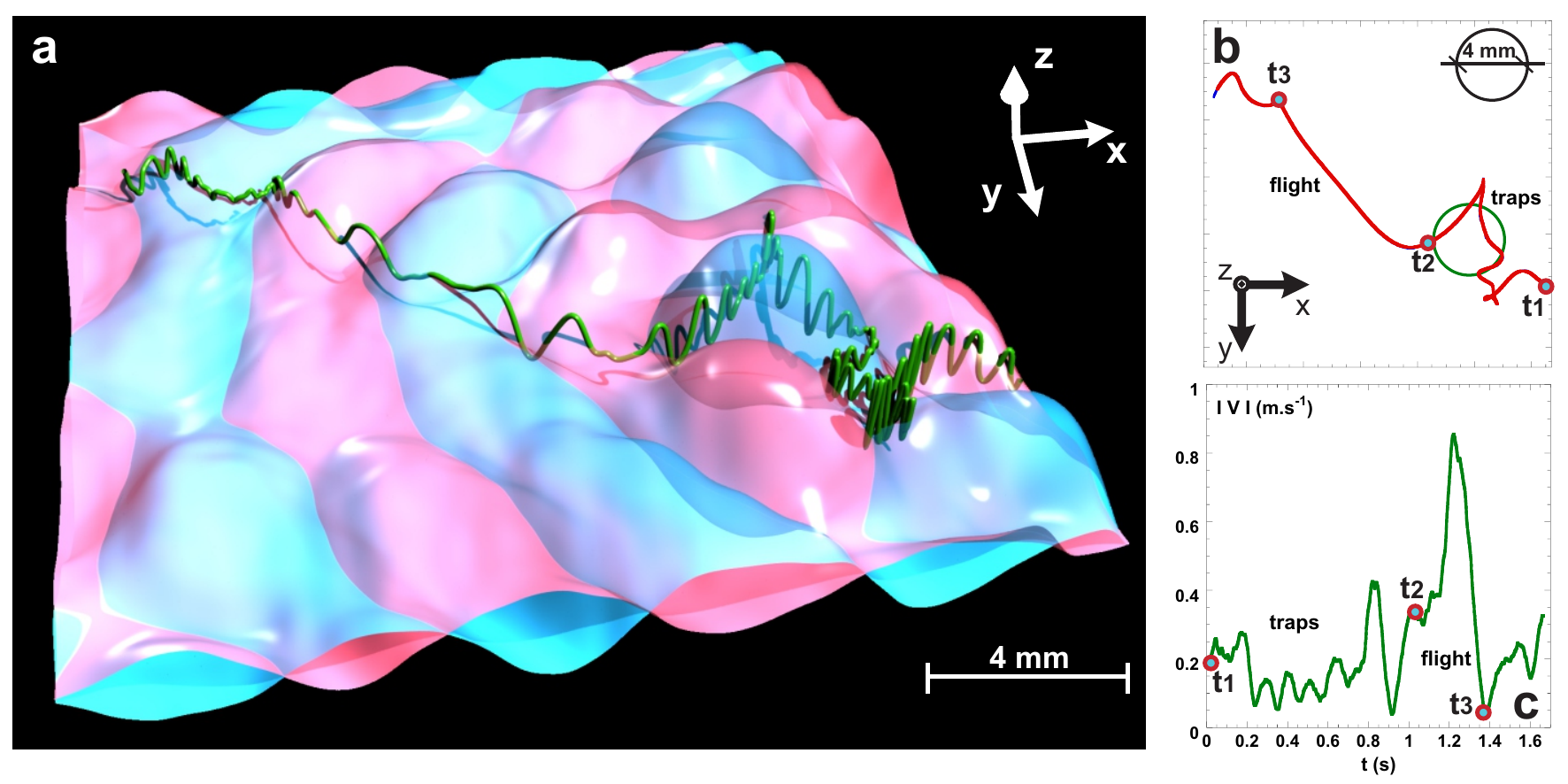}
\caption{\label{FigdisFW} (a) Perspective view of a three-dimensional particle trajectory followed for 1.6 s at a frame rate of 587 frames per second in a wave field parametrically excited at $a=1.6g$ ($a_{th}=0.6g$) and $f_0=60$ Hz. Pink and blue wave fields correspond to two consecutive phase extrema of the waves which are separated in time by one period of the shaker oscillation (Faraday waves are parametrically excited waves). (b) Horizontal 2D projection of the same trajectory. A green circle of radius $\lambda/4\approx2$ mm is indicated. (c) Absolute value of the horizontal particle velocity averaged over one Faraday period.}
\end{figure*}

In these experiments, the three-dimensional motion of floating tracers in Faraday waves and the topography of the wave field have been measured simultaneously.

Faraday waves are formed in a circular container (178 mm diameter, 30 mm deep) filled with water. The liquid depth is larger than the wavelength of the perturbations at the surface (deep water approximation). The container is vertically vibrated by an electrodynamic shaker. The forcing is monochromatic and is set to either $f_0=30$ Hz or $f_0=60$ Hz. Beyond a certain vertical acceleration threshold ($a>{a_{th}}$), the surface of the liquid becomes unstable. This is the Faraday instability. The waves that appear on the fluid surface, are parametrically forced; the dominant frequency of the excited surface ripples is at the first subharmonic of the excitation frequency, $f=f_0/2$. In the range of frequencies $f_0=\left(30-60\right)$ Hz, the Faraday wavelength is $\lambda\approx\left(8-14\right)$ mm. We define the supercriticality as $\epsilon=(a-a_{th})/{a_{th}}$ , where $a_{th}$ is the threshold for the parametric generation of Faraday waves.

The liquid surface is seeded with floating tracers to visualize the horizontal fluid motion. The use of surfactant and plasma treatment ensures that tracers do not aggregate (see the Appendix).

We use a diffusing light imaging technique to measure the topography of the wave field. A few percent of milk added to water provides sufficient contrast to compute high resolution reconstruction of the parametrically excited wave field (see the Appendix). Three-dimensional Lagrangian trajectories are obtained using a combination of two-dimensional PTV technique and a subsequent estimation of the local elevation along the trajectory (see the Appendix).

\subsection{3D features of Lagrangian trajectories in Faraday waves.}

 Figure~\ref{FigdisFW}(a) shows the 3D reconstruction of a trajectory recorded for 1.6 second in a disordered Faraday wave field forced at $f_0=60$ Hz and a vertical acceleration of $a=1.6g$ ($\epsilon\approx1.7$). This trajectory is superimposed on a reconstruction of the wave field which is built upon two images corresponding to two successive maxima of the parametric excitation (i.e: the blue and pink wave fields are separated by a time interval $dt=1/f_0=0.016$ s).
 
  Figure~\ref{FigdisFW} illustrates qualitatively two important features of the particle motion in disordered FW:\\
i) the particle motion can be extremely convoluted, often forming twisted patterns or even cusps which coincide with a strong orientational randomization of the trajectory (Fig.~\ref{FigdisFW}(b)). These intricate loops are usually confined within a circle whose radius is $\approx\lambda/4$. The lower the vertical acceleration, the longer the time a particle spends doing this complex motion. A pronounced cusp can be seen in Fig.~\ref{FigdisFW}(b). This is a singular point in the trajectory geometry which corresponds to a reversal of the particle velocity.\\
ii) a particle can escape this trapping and it experiences intense velocity peaks as shown in Fig.~\ref{FigdisFW}(c). From the top view (Fig.~\ref{FigdisFW}(b)), the particle trajectory during such event looks like a long ballistic flight. However, this ballistic behaviour is a characteristic of the horizontal motion only. Indeed, Fig.~\ref{FigdisFW}(a) clearly shows that the particle follows the complex \textit{z}-topography of the wave field during such a flight.\\
  In the following, we refer to 'traps' and 'flights' to characterize these geometrical features observed in all particle trajectories in disordered FW \cite{Hansen1997}. We also note that the particle trajectory shows poor resemblance to the classical Stokes drift picture \cite{Stokes1847}, where a trajectory is composed of slowly drifting vertical orbits, called trochoids.

 Following the approach described in \cite{Hansen1997}, we now give a quantitative definition of traps and flights along a trajectory. First, we compute the Lagrangian correlation time $T_{L}=\int^{\infty}_{0}\rho(t)dt$ where $\rho(t)$ is the temporal autocorrelation velocity function \cite{Taylor1921}, as well as the spatial Lagrangian correlation scale $L_{L}=\int^{\infty}_{0}\rho(r)dr$, where $\rho(r)$ is the spatial autocorrelation velocity function defined in \cite{XiaNC2013}. $L_{L}$ is the characteristic length scale of the particle dispersion. It was shown that $L_{L}\approx\lambda/2$ for developed 2D turbulence \cite{XiaNC2013,Xia2014}.

\begin{figure}[t!]
\includegraphics[width=6cm]{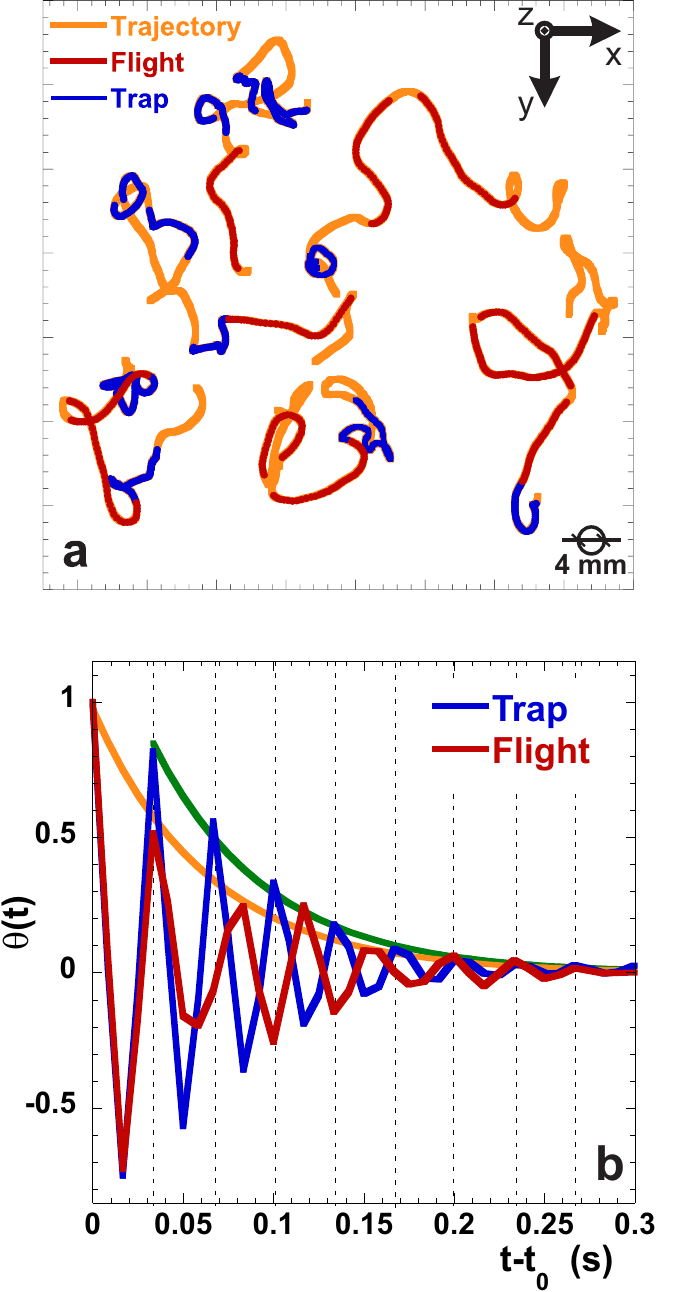}
\caption{\label{FigTF} (a) Detection of traps ($\delta r(t) < L_{L}$, blue segments) and flights ($\delta r(t) > L_{L}$, red segments) along particle trajectories (underlying orange lines). Traps and flights last at least $5T_{L}$ and $3T_{L}$ respectively. The forcing frequency is $f_0=60$ Hz, the Faraday wavelength is $\lambda\approx8$ mm and $T_{L}\approx0.10$ s. (b) Autocorrelation function $\theta(t)$ of the Lagrangian elevation estimated during traps and flights events for $a=1.6g$ and $f_0=60$ Hz. Exponentially decaying envelopes are plotted (green and orange lines). The dashed lines mark successive Faraday periods.}
\end{figure}

Then we consider, along the particle trajectory, the 'jumping function' defined as \cite{Hansen1997}:

\begin{equation}
\delta r(t)=\left|\stackrel{\rightarrow}{r}(t+T_{L}/2)-\stackrel{\rightarrow}{r}(t-T_{L}/2)\right|.
\end{equation}

A position $\stackrel{\rightarrow}{r}(t)$ is called stagnant if $\delta r(t) < L_{L}$, conversely if $\delta r(t) > L_{L}$ then $\stackrel{\rightarrow}{r}(t)$ is a jump. Finally we refer to as traps, any sequences of consecutive stagnant points which correspond to a time span larger than $5T_{L}$. On the other hand, a flight is a succession of jumps which lasts at least $3T_{L}$. Points located at the trajectory ends ($\delta t< T_{L}/2$) are not considered for the trap/flight analysis. Figure~\ref{FigTF}(a) shows the nature of traps and flights detected by these definitions. As expected, traps are characterized by extremely convoluted twists in all trajectories. On the other hand, flights correspond usually to the 'straightest' parts of the trajectories.

To assess correlations existing between the wave field topography and the horizontal features of the trajectory, the autocorrelation of the particle elevation was calculated in both traps and flights. It is defined as:

\begin{equation}
\left\langle \theta(t)\right\rangle=\langle[(z_{p}(t)z_{p}(t_{0})]\rangle,
\end{equation}

where $\left\langle \right\rangle$ denotes the statistical average, $z_{p}(t)$ is the particle elevation at time $t$ and $z_{p}(t_{0})$ is its initial elevation which is obtained either at the beginning of a trap or a flight.

We note that both autocorrelation functions in Fig.~\ref{FigTF}(b) show an exponentially decaying envelope with a decay time $T_{dec}$ longer than the Faraday period $T_{F}$. For traps we found $T_{dec}=2.9$ $T_{F}$, while $T_{dec}=1.9$ $T_{F}$ for the flights.
Figure~\ref{FigTF}(b) also shows that traps are clearly correlated with the subharmonic of the frequency excitation. This is because trapped particles rarely cross over the nodal lines around one wave peak. Conversely, the elevation autocorrelation function computed on flights reveals a substantial phase shift which can be interpreted as a Doppler shift. Indeed, in Fig.~\ref{FigdisFW}(a) the direction of the ballistic flight seems to be insensitive to the wave elevation and the flight itself corresponds to a 'straight' path probing different wave phases. In Fig.~\ref{FigdisFW}(a), we also note that a trap is clearly correlated with an area delimited by the nodal lines around a wave peak.

\subsection{2D Turbulence in 3D Faraday Waves.}

\begin{figure}[t!]
\includegraphics[width=9 cm]{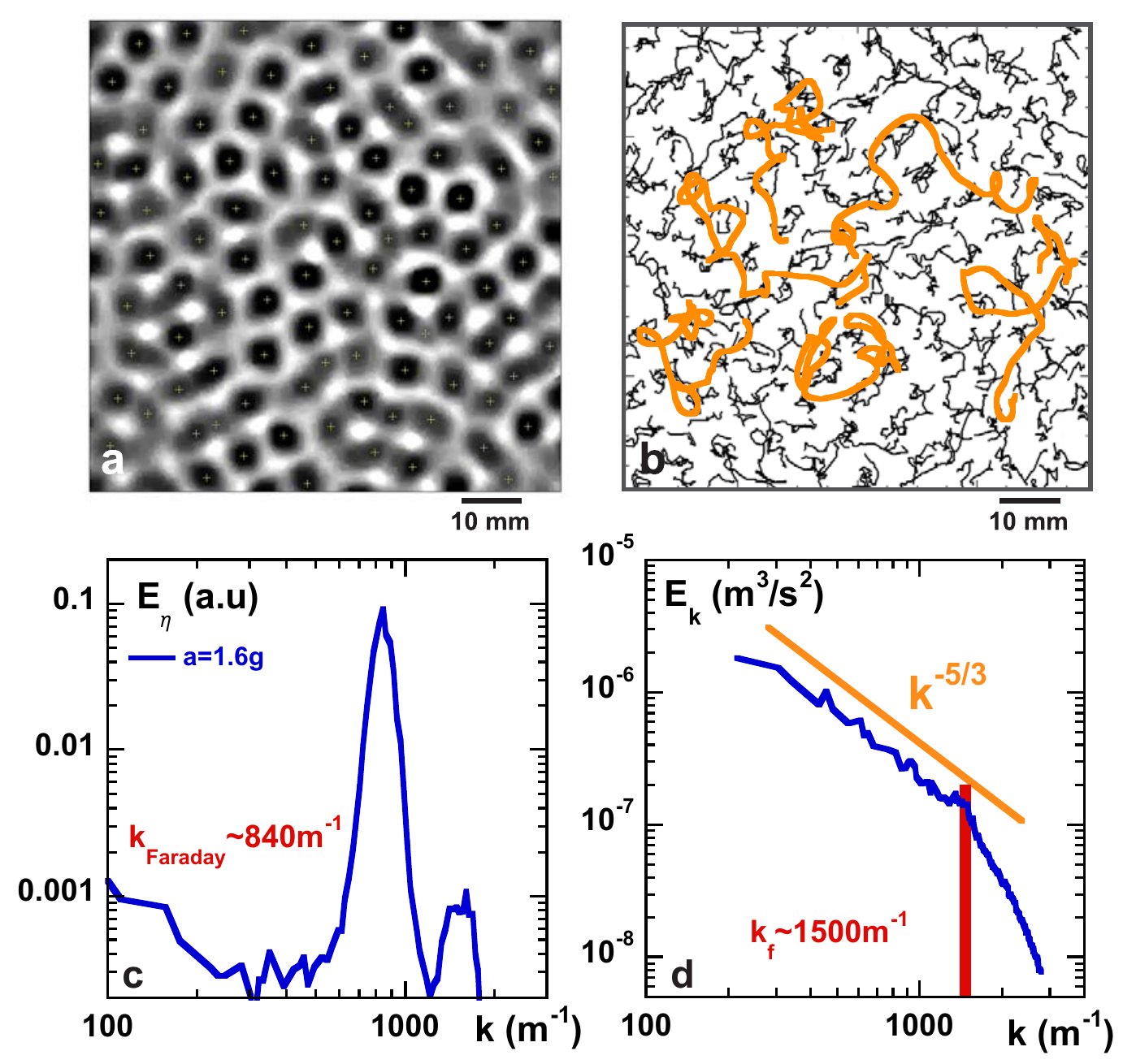}
\caption{\label{Fig4} In this Figure, FW are excited at $f_0=60$ Hz and $a=1.6g$ ($\epsilon=1.67$) at the water surface (water+10 $\%$ milk+surfactant) in a 178 mm diameter circular container. (a) Diffusing light image of the fluid surface elevation: peaks and troughs appear as dark and white blobs respectively. Particles have been removed from the raw image by using a rolling ball algorithm. Local wave maxima (dark peaks) are detected and subsequently followed by PTV techniques. (b) Comparison between trajectories of oscillons (black lines) having a phase maximum in (a) and an example of six particle trajectories  (orange lines). (c) Wave number spectrum of the surface elevation. (d) Wave number spectrum of the kinetic energy of the horizontal flow. The horizontal velocity field was measured using PIV techniques (see details in the Appendix). The red rectangle indicates a spectrally localized forcing mechanism with a characteristic length scale $l_{f}\approx4$ mm.}
\end{figure}

It has recently been shown that the Faraday wave properties are consistent with those of an ensemble of oscillating solitons, or oscillons \cite{Shats2012}. The oscillon peaks (Fig.~\ref{Fig4}(a)) can be tracked and these positions are used to construct oscillon trajectories. Figure~\ref{Fig4}(b) shows a visualization of these oscillon tracks along with six particle trajectories; both were followed over 4 s at a supercriticality of $\epsilon=1.67$. The oscillons and the particles motions occur on different length scales. An oscillon wanders erratically within a cage whose typical size is roughly $\lambda/2$ \cite{Xia2012}. Consequently, their confined tracks mark the sites of an irregular lattice, as confirmed by a peaked wave number spectrum of the surface elevation (see Fig.~\ref{Fig4}(c)). But particles move substantially in the wave field, similarly to what was observed in Fig.~\ref{FigdisFW}(a). Although the presence of these flights is connected with wave disorder \cite{Francois2013}, Fig.~\ref{Fig4}(b) emphasizes a clear distinction between the disorder-induced oscillon motion and the fluid motion.

Figure~\ref{FigdisFW}(c) suggests a possible connection between the particles horizontal motions and high velocity events, or flights. The key role played by inertia in particle behavior is even better illustrated in the Eulerian frame. Figure~\ref{Fig4}(d) shows a wave number spectrum of the horizontal kinetic energy of the fluid in Faraday waves. This spectrum was computed from Eulerian horizontal velocity fluctuations measured using PIV technique (see the Appendix) at $f_0=60$ Hz and a vertical acceleration $a=1.6g$ ($\epsilon=1.67$). Two scaling laws can be identified: $k^{-5/3}$ at wave numbers $k<1500$ m$^{-1}$ and $k^{-3}$ at $k>1500$ m$^{-1}$. As reported in \cite{Kameke2011,Francois2013}, these features are consistent with Kolmogorov-Kraichnan predictions for an inverse energy cascade and a direct enstrophy cascade respectively, which were derived for an incompressible fluid \cite{Kraichnan1967}. In these experiments, the vertical motion of the particles might result in non-zero values for the instantaneous divergence of the horizontal velocity field, however the divergence becomes close to zero when it is averaged over space or in time over several Faraday periods \cite{Kameke2011}. We also note that particles stay homogeneously distributed over the water surface. These observations confirm that the 2D flow generated by FW can be considered as incompressible. In contrast, floaters driven by an underlying 3D turbulence form massive ribbon-like clusters and their motion at the fluid surface exhibits properties of a compressible fluid \cite{Bandi2006,Larkin2009}.

 The kink observed in the spectrum of the horizontal kinetic energy at $k\approx1500$ m$^{-1}$ gives a forcing wavenumber corresponding roughly to half the Faraday wavelength $\lambda/2\approx4$ mm (see Fig.~\ref{Fig4}(c-d)). This observation raises the following question: how do Faraday waves generate a spectrally localized forcing mechanism of 2D turbulence?

\subsection{Wave induced vorticity.}

\begin{figure*}[!t]
\includegraphics[width=11 cm]{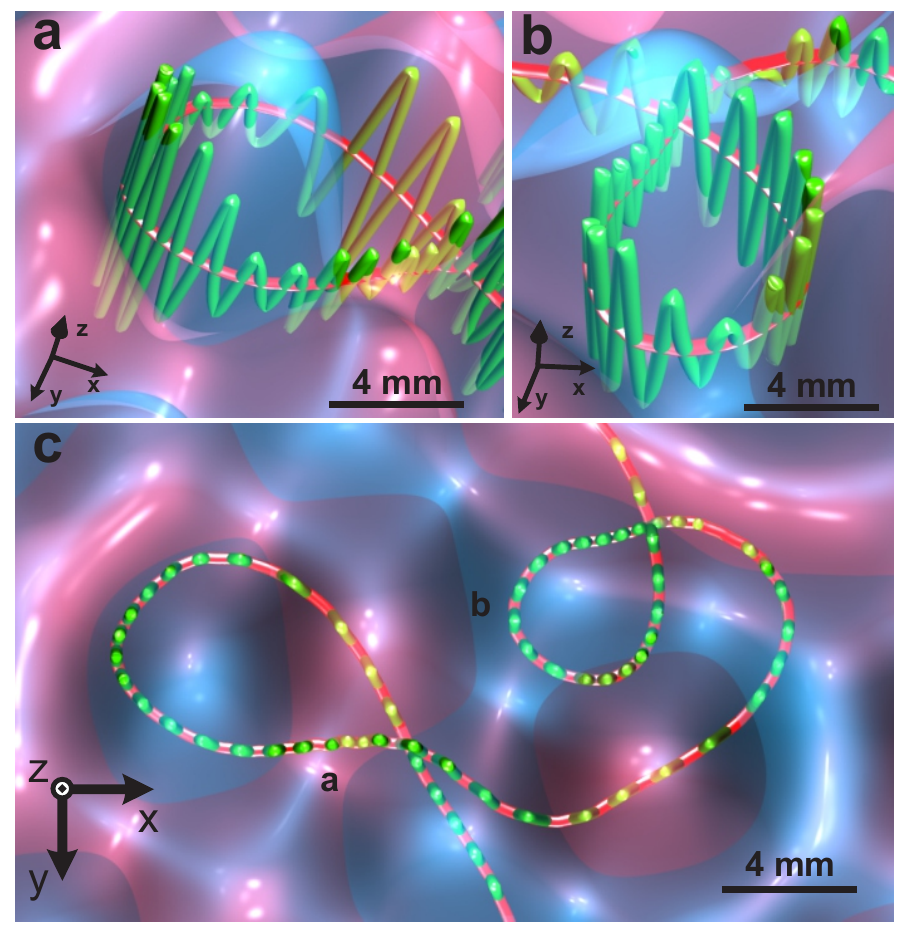}
\caption{\label{Figloop} Three-dimensional loops along one particle trajectory in FW at low drive $a=1g$ and $f_0=60Hz$ ($\epsilon=0.67$). The coordinate system indicates the different view points with z indicating the ascending vertical. The green line is the three-dimensional trajectory, the red line corresponds to the horizontal projection of this trajectory.  Pink and blue wave fields correspond to two consecutive phase extrema of the waves which are thus separated in time by one period of the shaker oscillation (Faraday waves are parametrically excited waves). (a-b) Close-up (perspective views) on almost perfectly circular loops along the trajectory, those loops have a radius close to $\approx\lambda/2$. (c) Top view (orthogonal projection) on a portion of this trajectory which corresponds to a 2 s interval. The view is zoomed out to show the two complete loops displayed in (a) and (b).}
\end{figure*}

\begin{figure*}[t!]
\includegraphics[width=16 cm]{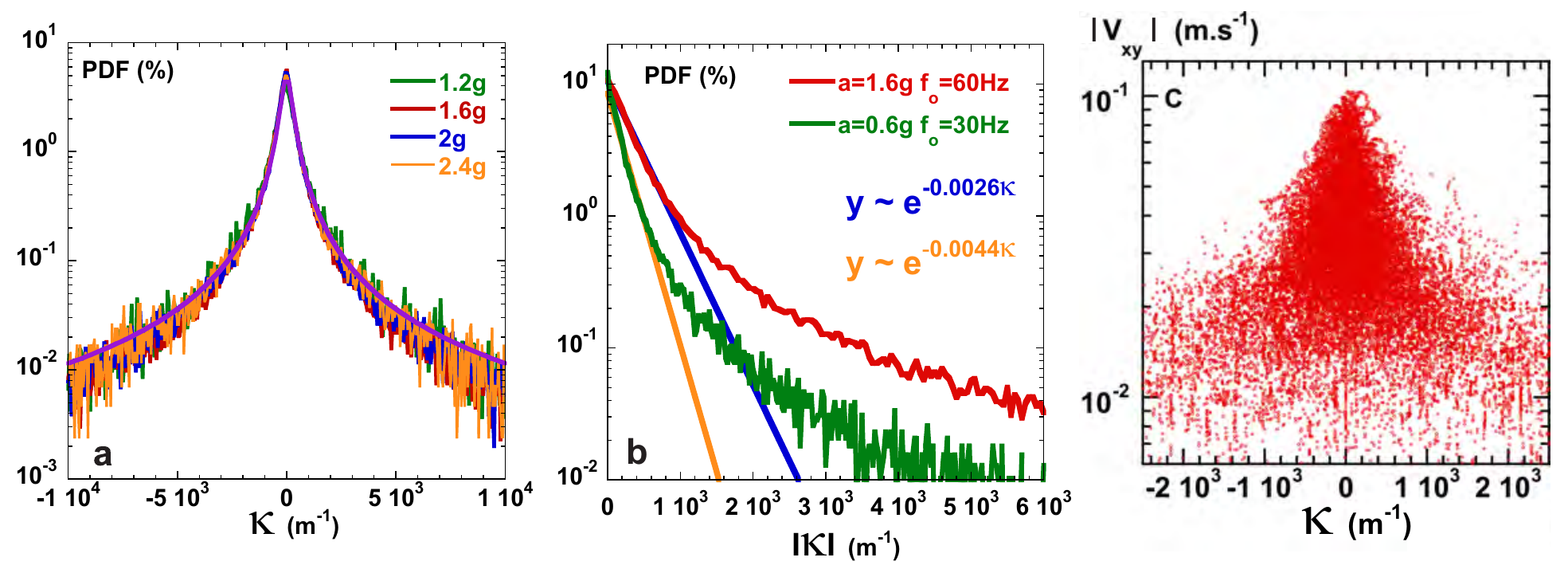}
\caption{\label{Figkappa} (a) PDFs of the curvature $\kappa_{T_{f}}(t)$ measured at $f_0=60$ Hz for different vertical accelerations $a=1.2,1.6,2,2.4$g ($a_{th}=0.6g$), the purple solid line corresponds to a fit by a $q$-gaussian distribution $P(\kappa)=\alpha\left[1-\beta(1-q).\kappa^{2}\right]^{1/1-q}$. (b) PDFs of the absolute value of the curvature $\left|\kappa_{T_{f}}(t)\right|$ for different excitation frequencies $f_0$=30, 60 Hz and similar supercriticality. The PDF at low curvature $\left|\kappa_{T_{f}}\right|<1000$ m$^{-1}$ is fitted by an exponential law $P(\kappa)=\alpha exp(-l_{c}\kappa)$ where $l_{c}$ is a characteristic decay length scale. The Faraday wavelength is $\lambda\approx14,8$ mm for $f_0$=30, 60 Hz respectively. (c) Absolute value of the horizontal velocity $\left|v_{xy}\right|$ versus curvature $\kappa_{T_{f}}(t)$ along trajectories recorded in FW at $f_0=60$ Hz ($a=2g$, $\epsilon=2.3$). Data plotted in Figures(a-c) were obtained from trajectories recorded in a $8\times8$ cm$^{2}$ domain imaged at $120$ Hz with a spatial resolution of 100 $\mu m$.}
\end{figure*}

 Figure~\ref{Figloop} shows details of a three dimensional trajectory followed for 2 seconds in FW at relatively low drive ($\epsilon=0.67$). By decreasing the vertical acceleration, the average 'trapping time' increases \cite{Hansen1997}. Thus, some trajectories can consist almost exclusively of consecutive trapping sequences as seen in Fig.~\ref{Figloop}(c). Figure~\ref{Figloop} also reveals an interesting feature of particle trajectories that is somewhat hidden in the complex Lagrangian picture of Figure~\ref{FigdisFW}(a): traps can look like almost perfectly circular loops around an oscillon. It suggests that waves may generate local vorticity in the horizontal flow. In Figures~\ref{Figloop}(a-c), these horizontal loops have a diameter comparable to the size of an oscillon ($\approx\lambda/2$).

 Moreover, Figure~\ref{Figloop} emphasizes that this circular particle motion at the oscillon scale is somehow independent of the Faraday period but seems related to a much longer time scale. In Fig.~\ref{Figloop}(a) for instance, a particle describes an almost circular loop which takes 24 Faraday periods corresponding to the peaks in the green trajectory.
 
 We now consider the horizontal projection of the particle trajectory and test if there is any resilient feature of those circular loops when the vertical acceleration is increased. The geometry of the horizontal projection of particle trajectory can be characterized by its local curvature $\kappa$. This quantity reveals the amount by which a trajectory deviates from a straight line and thus indicate the presence of loops or cusps: low values of $\kappa$ are related to locally straight flights while higher values of $\kappa$ occur in traps. The instantaneous curvature $\kappa(t)$ along a trajectory is given by:

\begin{equation}
\kappa(t)\stackrel{\rightarrow}{z}=\left[\stackrel{\rightarrow}{v}_{xy}(t)\times\stackrel{\rightarrow}{a}_{xy}(t)\right]/\left|\stackrel{\rightarrow}{v}_{xy}(t)\right|^{3}
\end{equation}

where $\stackrel{\rightarrow}{v}_{xy}$ and $\stackrel{\rightarrow}{a}_{xy}$ stand for the horizontal velocity and acceleration along a trajectory, $\stackrel{\rightarrow}{z}$ is the unit vector in the vertical direction. As the computation of $\stackrel{\rightarrow}{a}_{xy}$ is quite sensitive to the noise level, the curvature $\kappa(t)$ was smoothed over one Faraday period $T_{f}$, such that:

\begin{equation}
\kappa_{T_{f}}(t)=\frac{1}{T_{f}}\int^{t+T_{f}}_{t}\left[\kappa(t)\right]dt
\end{equation}

On a technical note, the local curvature diverges when the velocity $\stackrel{\rightarrow}{v}_{xy}$ is zero (i.e. the displacement is smaller than the spatial resolution which is 100 $\mu m$ for the data plotted in Fig.~\ref{Figkappa}). These events correspond to the presence of sharp cusps along the trajectory (see Fig.~\ref{FigdisFW}(b)). In such rare cases, locally diverging points in $\kappa(t)$ are removed before the quantity is smoothed over $T_{f}$.

In Figure~\ref{Figkappa}, we show the probability density function (PDF) of the curvature for different excitation frequencies $f_0$ and vertical accelerations. First, these PDFs are wide with distinct tails.
The PDFs are found independent of the vertical acceleration for a fixed $f_0$, Fig.~\ref{Figkappa}(a). However, the PDF becomes more peaked and narrows around the origin with the decrease in the forcing frequency $f_0$. It keeps a power law tail but high $\kappa_{T_{f}}$ events get less probable, Fig.~\ref{Figkappa}(b). Although the entire PDF follows a $q$-gaussian distribution, it can be fitted by an exponential decay law $P(\kappa_{T_{f}})=\alpha$exp($-l_{c} \kappa_{T_{f}}$) for the curvature values within the range $\left|\kappa_{T_{f}}\right|\leq1000$ m$^{-1}$. Since the curvature is the inverse of the curvature radius, $l_{c}$ has the dimension of a length scale. This characteristic length scale is roughly a quarter of the Faraday wavelength as seen from the fitting parameters for two different cases at $f_0=30$ Hz and $f_0=60$ Hz in Figure~\ref{Figkappa}(b).

In Figure~\ref{Figkappa}(c), the absolute value of the horizontal velocity $\left|\stackrel{\rightarrow}{v}_{xy}(t)\right|$ is compared with the curvature $\kappa_{T_{f}}(t)$ in disordered FW ($a=1.6g$, $\epsilon=1.67$). High values of $\left|\stackrel{\rightarrow}{v}_{xy}\right|$ are correlated with low values for $\kappa_{T_{f}}$. Thus fluid parcels with high kinetic energy follow straight ballistic flights. On the other hand, high curvature points, which characterize loops and cusps, are related to low velocities. Thus slow particles are trapped along convoluted trajectories around oscillon sites.

\begin{figure*}[t!]
\includegraphics[width=13 cm]{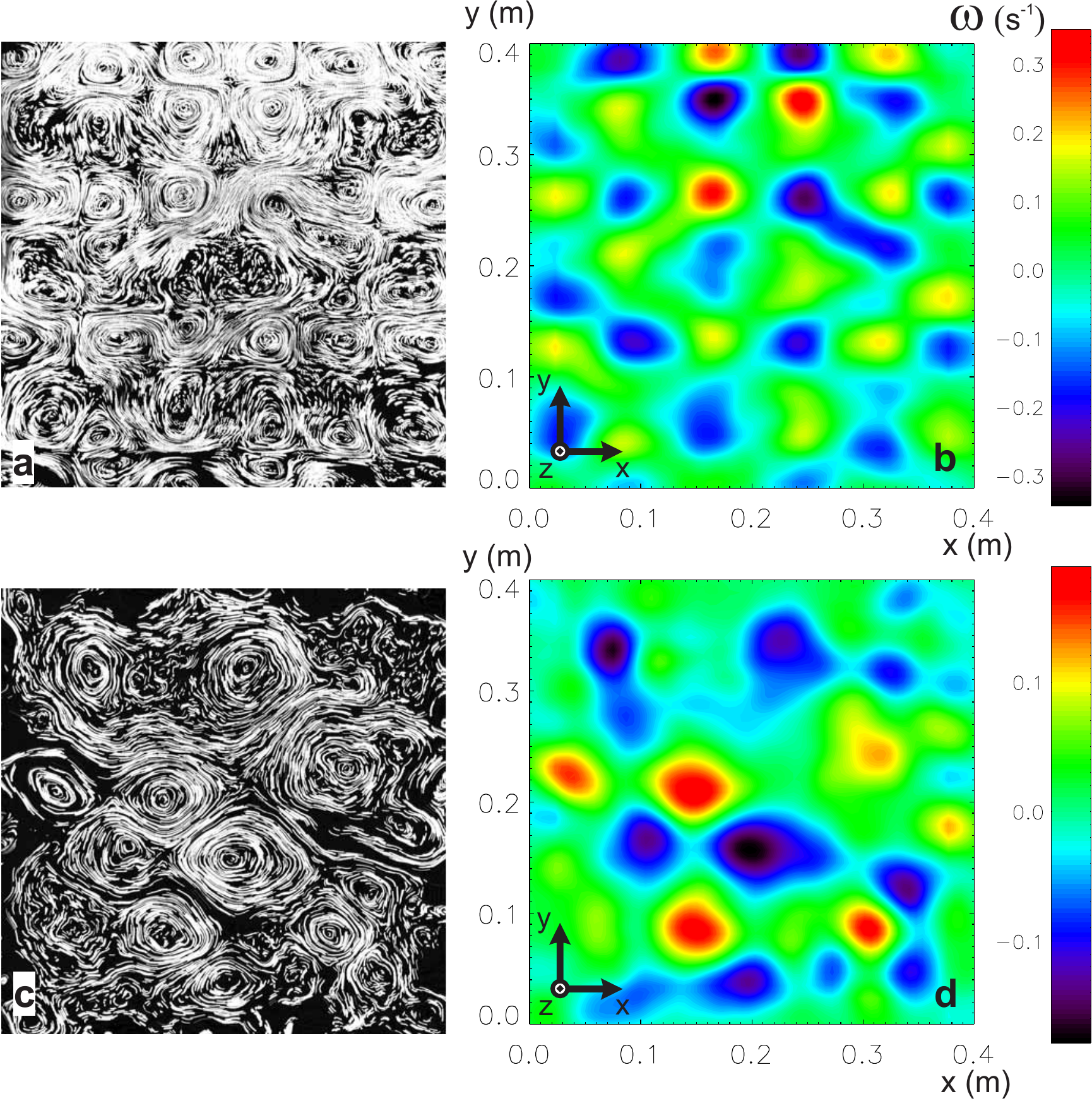}
\caption{\label{FigVortarray} Flow driven by Faraday waves excited at the water surface at $f_0=10$ Hz and $a=0.04$g in a $40\times40 cm^{2}$ square container. (Left) Fluid particle trajectories 1 minute (a), 10 minutes (c) after the shaker has been turned on. In those images, particle streaks correspond to a 4 s time interval. (Right) Corresponding vorticity fields measured by PIV techniques (see details in the Appendix) and averaged over a 4 s time interval ($=20$ $T_{f}$). Movies were recorded at 5 fps with a spatial resolution of 200 $\mu$m.}
\end{figure*}

The relation between the Lagrangian curvature and the Eulerian vorticity is far from being trivial. 
To ascertain the claim that FW actually generate vorticity in the horizontal flow and to clarify further the existence of a unique scale for the vorticity injection, we have performed an additional set of experiments. Here Faraday waves are produced in a large ($400$ mm) square container. At the excitation frequency of $f_0$=10 Hz, a quasi-regular square lattice of $7\times7$ oscillons is excited.
  Shortly after the parametric waves are generated, a matrix of vortices is formed which roughly corresponds to the underlying quasi-regular lattice of oscillons Fig.~\ref{FigVortarray}(a). The characteristic diameter of the vortices is close to the size of an oscillon. This configuration however is not stable as the vortices interact with each other. The flow evolves as illustrated in  Figure~\ref{FigVortarray}(c), the vortex interaction results in the emergence of a disordered flow.

  The evolution of the vertical vorticity field $\stackrel{\rightarrow}{\omega}_{z} =(\delta u_y/\delta x-\delta u_x/\delta y)\stackrel{\rightarrow}{z}$ computed using PIV velocity field (see details in the Appendix) and averaged over 4 seconds is shown in Fig.~\ref{FigVortarray}(b) and (d). In Fig.~\ref{FigVortarray}(b), the vorticity field consists of an almost regular square pattern made of counter-rotating vortices. These vortices ultimately aggregate into larger eddies (Fig.~\ref{FigVortarray}(d)). The total vorticity remains zero. 

\newpage

\section{III. DISCUSSION and CONCLUSIONS}

  \textit{How do Faraday waves generate a spectrally localized forcing mechanism of 2D turbulence?} 
 
  Figure \ref{FigVortarray} demonstrates that Faraday waves create horizontal vortices on the liquid surface at the early stage of the flow development. Initially a periodic array of vortices can be detected. These vortices have a diameter given by the size of an oscillon. As the flow develops, these vortices merge into larger eddies in the process of the inverse energy cascade. The initial vortices are the fuel of 2D turbulence.
 
  The creation of vorticity by Faraday waves bears striking similarity with that in electromagnetically driven turbulence. Though the forcing scale vortices in electromagnetic turbulence are generated by periodic arrays of permanent magnets, the electromagnetically driven vortices injected into the flow are very similar to those illustrated in Fig.\ref{FigVortarray}(a). This explains why the 2D turbulence produced by these two different methods is very similar \cite{XiaNC2013}. In both methods, the forcing scale vortices are only visible in the Eulerian frame in the initial stage of the turbulence development \cite{Shats2005}.
  
  The experimental data show that the shape of the curvature PDF is a resilient feature of particles trajectories in disordered FW. The most probable values for the curvature can be fitted by an exponential decay law. The characteristic decay diameter is determined by the period of the oscillon lattice \cite{Xia2012,Francois2013} which is also the forcing scale of the kinetic energy spectrum, Fig.~\ref{Fig4}(d). We note that a connection between a statistical property of the Lagrangian curvature ($l_{c}$ in Fig~\ref{Figkappa}(b)) and the forcing scale detected in the kinetic energy spectrum (Fig.~\ref{Fig4}(d)) is not trivial. Indeed the Lagrangian and Eulerian descriptions of a flow are difficult to connect. Hence it is remarkable that the curvature of the Lagrangian trajectories still reveals a "footprint" of the forcing scale vortices in fully developed turbulence.\\

\begin{figure}[!b]
\includegraphics[width=7 cm]{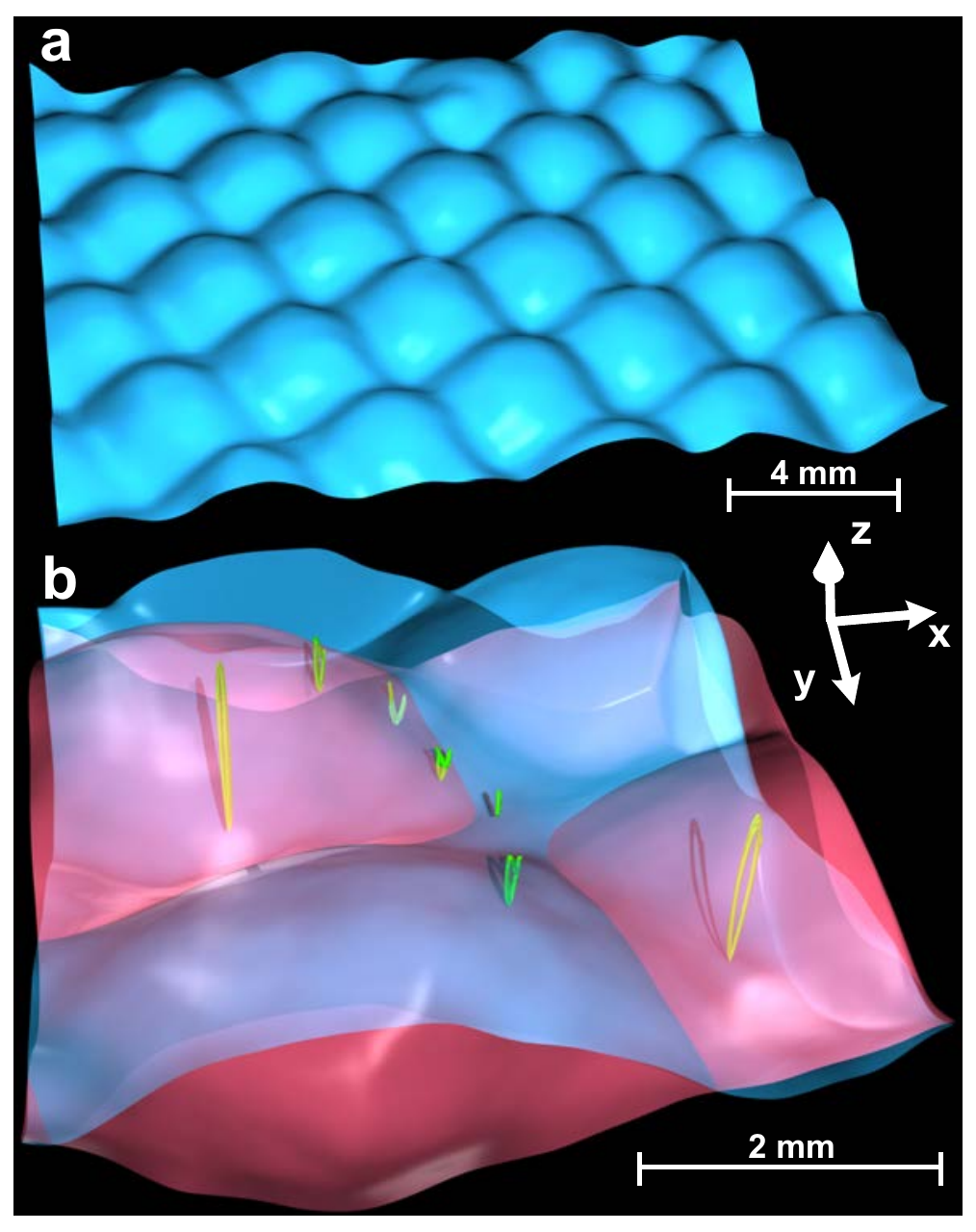}
\caption{\label{FigcryFW} Perspective view of three-dimensional particle trajectories in a FW crystal. FW are excited at the surface of a solution of $73\%$ glycerol + $10\%$ milk + water (no added surfactant). The vertical acceleration is $a=2.8g$ ($a_{th}=2.6g$, $\epsilon=0.08$) at $f_0=60$ Hz. (a) Waves showing a crystalline pattern. (b) Close-up of 3D particle trajectories followed for 8 Faraday periods ($8T_{f}\approx 264$ ms) at the wavelength scale. Pink and blue wave fields correspond to two consecutive phase extrema of the waves which are thus separated in time by one period of the shaker oscillation (Faraday waves are parametrically excited waves).}
\end{figure}

\textit{The Lagrangian motion in a 3D Faraday wave crystal.}

 Figure~\ref{FigcryFW} is to be compared with Figure~\ref{FigdisFW}. It shows the particle trajectories observed in a crystal-like wave field. An almost perfect wave crystal is realized in a viscous solution of glycerol ($73\%$), milk ($10\%$), and water (viscosity = 40 cP) \cite{Fauve1994}.
In such a crystal-like array of waves, Fig.~\ref{FigcryFW}(a), particles no longer experience spatially extended random walks. Instead, they move along closed orbits of different geometries, Fig.~\ref{FigcryFW}(b). The orbital motion is almost vertical near a wave antinode while it shows small horizontal oscillations close to a nodal line. Along a nodal line, particle trajectories are U-shaped. The cross-over of two nodal lines marks the location of a saddle point in the wave topography where there is no particle motion.

As the vertical acceleration is increased (or if dissipation is decreased), this ordered wave structure breaks down into a disordered lattice of steep waves. Although Faraday waves remain almost standing waves, the presence of disorder in the wave field coincides with an increase in the particle mobility and the emergence of 2D Navier-Stokes turbulence. These observations suggest that 2D turbulence should be considered as a potential source of disorder in FW.\\

\textit{Conclusions.}

These experiments prove that the deformation of a fluid interface can be considered as a new paradigm for vorticity creation in a 2D flow \cite{Saffman}. Besides, this finding sheds new light on the wave induced fluid motion which is commonly considered in textbooks as an irrotational flow. As a consequence, the influence of vorticity in surface wave phenomena was only studied relatively recently compared to the long standing history of surface wave physics. To our knowledge, the influence of vorticity has been mainly addressed in viscous boundary layers \cite{L-Higgins1953,L-Higgins1960}, or in coastal shear flows in oceanography \cite{Shrira1996}. In both cases, the presence of vorticity can have significant effects on the flow. In the latter case, the presence of strong nonlinearities in the governing equations has already stressed the importance of interacting vortices.

In Faraday ripples, a new physical model emerges: a 2D lattice of oscillating solitons contains vertical oscillatory energy, but it also produces horizontal vortices at the oscillon size: the 'turbulent fuel'. This horizontal energy is ultimately spread over a wide range of scales by the inverse energy cascade. This unexpected generation of 2D turbulence is an interesting twist in our understanding of the order-disorder transition in Faraday waves \cite{Gollub1989,Ciliberto1984,Shani2010}.

\newpage

\section{APPENDIX: MATERIALS AND METHODS}

\begin{figure*}[!t]
\includegraphics[width=18 cm]{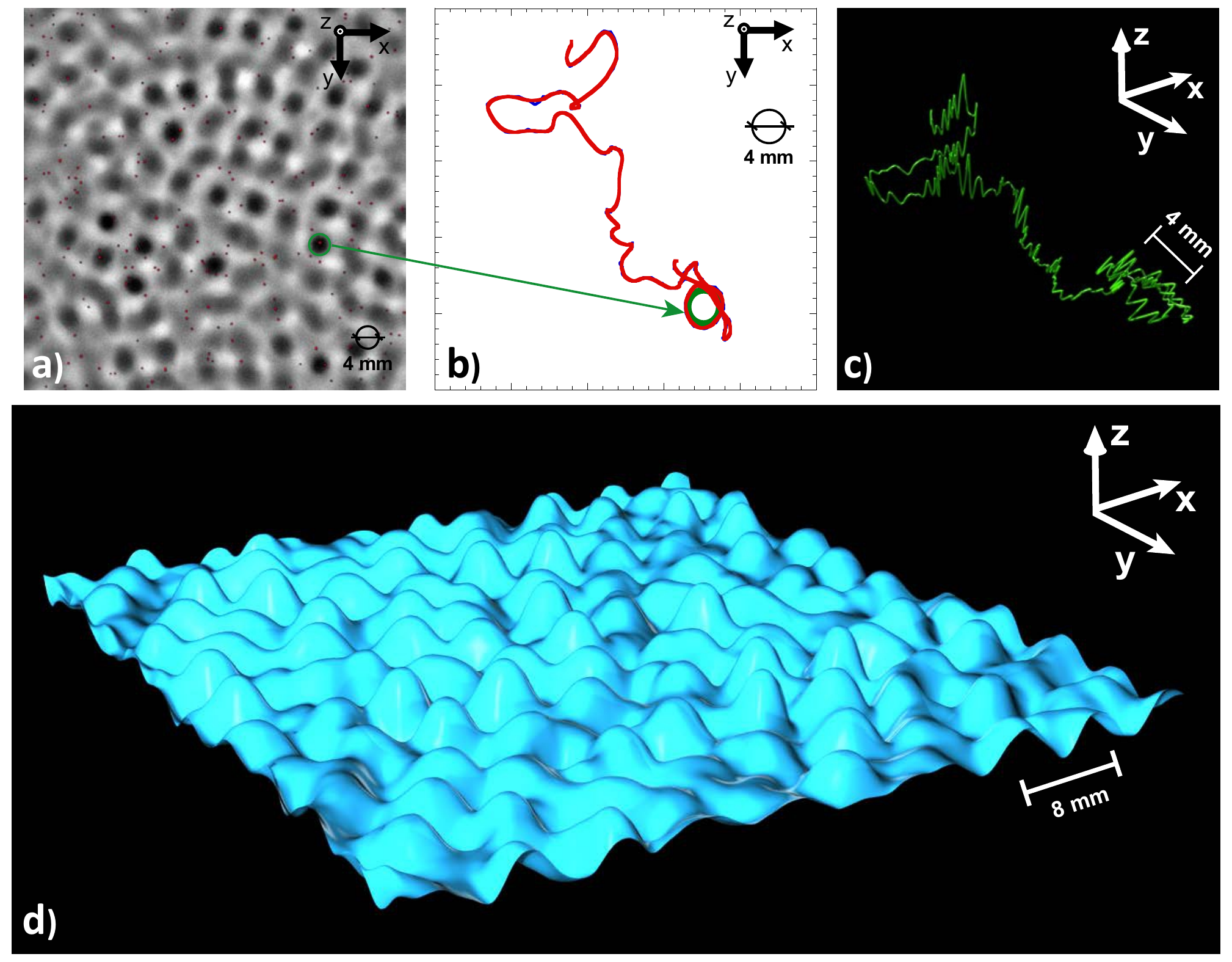}
\caption{\label{Figsetup} (a) Diffusing light image of the wave field at $a=2g$ ($\epsilon=2.3$) and $f_0=60$ Hz. Dark and light blobs give the location of wave peaks and troughs respectively. The tiny red dots are tracer particles (diameter=150-300 $\mu m$), the locations of which are extracted from the wave field using a 'rolling ball' filter. A green circle of radius $\lambda/4=2$ mm is indicated. (b) Slight zoom-in view on the horizontal projection of an individual trajectory followed for 4 s at 120 frames per second. This trajectory is reconstructed using PTV technique. The blue line (barely visible) corresponds to raw data points, the red line is the trajectory smoothed over one Faraday period ($T=1/f \approx 33$ ms). (c) 3D reconstruction of the trajectory shown in b). (d) 3D Visualization of the wave field topography shown in a). Particles have been removed from the image by using a 'rolling ball' filter, the original image ($80\times80$ mm$^{2}$) was smoothed over a $0.8\times0.8$ mm$^{2}$ window.}
\end{figure*}

Faraday waves are formed in a circular container (178 mm diameter, 30 mm deep) filled with a liquid whose depth is larger than the wavelength of the perturbations at the surface (deep water approximation). The container is vertically vibrated by an electrodynamic shaker (Bruel\&Kjaer). The forcing is monochromatic and set to either $f_0=30$ Hz or $f_0=60$ Hz. The amplitude $a$ of the vertical acceleration imposed by the shaker is measured by an accelerometer. An accurate control of the acceleration amplitude is performed by a proportional-integral-derivative controller.

We use a diffusing light imaging technique to measure the topography of the wave field, see Fig.~\ref{Figsetup}(a) and ~\ref{Figsetup}(d). The fluid surface is illuminated by a LED panel placed underneath the transparent bottom of the container. A few percent of milk (from $2\%$ to $10\%$) added to water provides sufficient contrast to obtain a high resolution reconstruction of the parametrically excited wave field. The absorption coefficient is calibrated before each experiment, by gradually increasing the liquid depth and by measuring the change in the light intensity transmitted when the liquid surface is flat (i.e. when the container is not vibrated). This procedure allows to calibrate the wave field elevation in millimeters. In these experiments, the dynamic range of our images (16 bits) allows to resolve a 20 $\mu m$ change in the fluid elevation.

Videos are recorded at a 16 bit resolution and at a high frame rate using the Andor Neo sCMOS camera which was mounted above the tank. The typical field of view is either a $8\times8$ cm$^{2}$ domain, imaged at $120$ Hz with a resolution of 100 $\mu m$ or a $3\times3$ cm$^{2}$ area, imaged at $587$ Hz with a resolution of 200 $\mu m$. Black floating particles spread on the fluid surface are easy to observe and allow to visualize the horizontal motion of the fluid. We use particles with a diameter within a range of 150-300 $\mu m$ (300 $\mu m$ is the mesh size of the sieve used). Particles are made of carbon glass and have been plasma treated to reduce their intrinsic hydrophobicity. The use of surfactant and plasma treatment ensures that particles do not aggregate on the surface. This is illustrated in Fig.~\ref{Figsetup}(a) with particles homogeneously distributed at the surface of the wave field.

To test if the particle size affects the observation of turbulence generation in Faraday waves, additional measurements were performed using 50 $\mu m$ diameter polyamid particles. No effect of the particle size was detected, in agreement with observations reported in  \cite{Kameke2011,XiaNC2013}.

\textit{3D-PTV techniques:}

Three-dimensional Lagrangian trajectories are retraced using a combination of two-dimensional PTV technique and a subsequent estimation of the local elevation along the trajectory (see Fig.~\ref{Figsetup}(b) and (c)). First, the horizontal projection (\textit{x-y} coordinates) of each point on a trajectory is tracked using a nearest neighbor algorithm \cite{Crocker1996}. In such algorithm, the maximum displacement allowed for a particle in consecutive frames is set to be smaller than the minimal distance separating particle pairs in the field of view. Then the particle elevation (\textit{z} coordinate) is estimated as the mean of the wave elevation over a local window (400 $\mu m$ radius) which is centered on the \textit{x-y} particle coordinates at a given time. In all our experiments, the localness of the \textit{z} coordinate estimation is ensured by choosing a sufficiently large Faraday wavelength (at $f_0=60$ Hz, $\lambda>>400$ $\mu m$). For each vertical acceleration \textit{a} at a given forcing frequency $f_0$, thousands of particle trajectories were followed for 4 s at 120 or 587 frames per second. The 3D trajectories of the particle and the wave field are visualized using the Houdini$^{(TM)}$ 3D animation tools (by Side Effects Software).
For the study of the horizontal features of particle trajectories, \textit{x-y} particle coordinates are first smoothed over one Faraday period $T_{f}$. This filtering limits the noise induced by the projection on the horizontal plane of the vertical oscillatory motion. Similarly, the determination of the curvature $\kappa(t)$ requires the computation of the horizontal acceleration $\stackrel{\rightarrow}{a}_{xy}$, a quantity quite sensitive to the noise level. To reduce the noise, the measured curvature $\kappa(t)$ was also smoothed over one Faraday period $T_{f}.$

\textit{2D-PIV techniques:}

The particle image velocimetry (PIV) technique is used to obtain the velocity field of the horizontal motion of the flow in Figure~\ref{Fig4}(d) and Figures~\ref{FigVortarray}(b) and (d). Images are split into a large number of interrogation windows which form a regular grid. A velocity vector is calculated for each interrogation area by temporal cross-correlation of the particle intensity distribution within each interrogation tile. The PIV technique allows to resolve sub-pixel displacement and thus gives a high degree of accuracy in the estimation of the velocity field.

In Figure~\ref{Fig4}(d), the flow is recorded at a frame rate twice the shaker frequency (i.e. equal to four times the parametric wave frequency) for 4 seconds. The field of view is $80\times80$ mm$^{2}$ with a spatial resolution of 100 $\mu m$ (pixel size). The PIV velocity fields are computed on a 90$\times$90 spatial grid (grid mesh size is 0.89 mm), with a $2.6\times2.6$ mm$^{2}$ interrogation window size  (the interrogation windows are overlapping). The measurements of the instantaneous displacement are accurate down to 10 $\mu m$. The energy spectrum in Figure~\ref{Fig4}(d) is averaged over 400 snapshots of the velocity field.

In Figures~\ref{FigVortarray}(b) and (d), the flow is recorded at a frame rate of 5 fps for more than 10 minutes. The field of view is $400\times400$ mm$^{2}$ with a spatial resolution of 200 $\mu m$ (pixel size). The PIV velocity fields are computed on a 70$\times$70 spatial grid (grid mesh size is 5.7 mm), with a $10\times10$ mm$^{2}$ interrogation window size (the interrogation windows are overlapping). The measurements of the instantaneous displacement are accurate down to 20 $\mu m$. 
The vertical vorticity fields in Figures~\ref{FigVortarray}(b) and (d) are averaged over a 4 second time interval ($=20$ $T_{f}$). The average vorticity measurement is accurate down to $10^{-3}$ $s^{-1}$.

\newpage

\section{Acknowledgments.}This work was supported by the Australian Research Council's Discovery Projects funding scheme (DP110101525). HX would like to acknowledge the support by the Australian Research Council's Discovery Early Career Research Award (DE120100364).


\begin{thebibliography}{02}

\bibitem{Faraday1831} M. Faraday, \textit{On the forms and states assumed by fluids in contact with vibrating elastic
surfaces}, Philos. Trans. R. Soc. London \textbf{121}, 319 (1831).
\bibitem{Krutsenko1978} I. V. Krutsenko, V. S. L'vov, G. A. Melkov, Zh. Eksp. Teor.Fiz. \textbf{75}, 1114 (1978) [\textit{Spectral Density of Parametrically Excited Waves}, Sov. Phys. JETP \textbf{48}, 561 (1978)].
\bibitem{Rinberg1997} D. Rinberg, V. Cherepanov, and V. Steinberg, \textit{Universal Broadening of Frequency Spectra in Parametrically Driven Systems}, Phys. Rev. Lett. \textbf{78}, 4383 (1997).
\bibitem{Rinberg2001} D. Rinberg, and V. Steinberg, \textit{Parametric Generation of Second Sound in Superfluid Helium: Linear Stability and Nonlinear Dynamics}, Phys. Rev. B \textbf{64}, 054506 (2001).
\bibitem{Nishikawa1968} K. Nishikawa, \textit{Parametric Excitation of Coupled Waves. II. Parametric Plasmon-Photon Interaction}, J. Phys. Soc. Jpn. \textbf{24}, 1152 (1968).
\bibitem{Rodishevskii1988} A. V. Rodishevskii, V. V. Savel'ev, V. B. Cherepanov, Zh. Prikl. Mekh. Tekh. Fiz. \textbf{4}, 61 (1988) [\textit{Threshold of Parametric Excitation of Waves on a Fluid Surface}, J. Appl. Mech. Tech. Phys. \textbf{29}, 512 (1989)].
\bibitem{Holt1996} R. G. Holt, and E. H. Trinh, \textit{Faraday Wave Turbulence on a Spherical Liquid Shell}, Phys. Rev. Lett. \textbf{77},1274 (1996).
\bibitem{Falcon2009} C. Falcon, E. Falcon, U. Bortolozzo, and S. Fauve, \textit{Capillary Wave Turbulence on a Spherical Fluid Surface in Low Gravity }, Europhys. Lett. \textbf{86},14002 (2009).
\bibitem{Brazhnikov2002} M. Yu. Brazhnikov et al., \textit{Observation of Capillary Turbulence on the Water Surface in a Wide Range of Frequencies}, Europhys. Lett. \textbf{58}, 510 (2002).
\bibitem{XiaShats2012} H. Xia and M. Shats, \textit{Propagating Solitons Generated by Localized Perturbations on the Surface of Deep Water}, Phys. Rev. E \textbf{85}, 026313 (2012).
\bibitem{Cross1993} M.C. Cross and P.C. Hohenberg, \textit{Pattern Formation Outside of Equilibrium}, Rev. Mod. Phys. \textbf{65}, 851 (1993).
\bibitem{Douady1990} S. Douady, \textit{Experimental Study of the Faraday Instability}, J. Fluid Mech. \textbf{221}, 383 (1990).
\bibitem{Gluckman1995} B. J. Gluckman, C. B. Arnold, and J. P. Gollub, \textit{Statistical Studies of Chaotic Wave Patterns}, Phys. Rev. E \textbf{51}, 1128 (1995).
\bibitem{Kudrolli1996} A.Kudrolli and J.P.Gollub, \textit{Patterns and Spatiotemporal Chaos in Parametrically Forced Surface Waves: Systematic Survey at Large Aspect Ratio}, Physica D \textbf{97}, 133 (1996).
\bibitem{Kityk2005} A.V. Kityk, J. Embs, V.V. Mekhonoshin, and C. Wagner, \textit{Spatiotemporal Characterization of Interfacial Faraday Waves by Means of a Light Absorption Technique}, Phys. Rev. E \textbf {72}, 036209 (2005).
\bibitem{Fauve1994} W.S.Edwards, and S.Fauve, \textit{Patterns and Quasi-patterns in the Faraday Experiment}, Journal of Fluid Mechanics \textbf{278}, 123 (1994).
\bibitem{Umbanhowar1996} P.B. Umbanhowar, F. Melo, and H.L. Swinney, \textit{Localized Excitations in a Vertically Vibrated Granular Layer}, Nature \textbf{382}, 793 (1996).
\bibitem{Goldman2003} D.I. Goldman, M.D. Shattuck, Sung Joon Moon, J.B. Swift, and H.L. Swinney, \textit{Lattice Dynamics and Melting of a Nonequilibrium Pattern}, Phys. Rev. Lett. \textbf{90}, 104302 (2003).
\bibitem{Gollub1989} N.B.Tufillaro, R.Ramshankar, and J.P.Gollub, \textit{Order-Disorder Transition in Capillary Ripples}, Phys. Rev. Lett. \textbf{62}, 422 (1989).
\bibitem{Ciliberto1984} S.Ciliberto, and J.P.Gollub, \textit{Pattern Competition Leads to Chaos}, Phys. Rev. Lett. \textbf{52}, 922 (1984).
\bibitem{Shani2010} I. Shani, G. Cohen, and J. Fineberg, \textit{Localized Instability on the Route to Disorder in Faraday Waves}, Phys. Rev. Lett. \textbf{104}, 184507 (2010).
\bibitem{Fineberg2000} H.Arbell, and J.Fineberg, \textit{Temporally Harmonic Oscillons in Newtonian Fluids}, Phys. Rev. Lett. \textbf{85}, 756 (2000).
\bibitem{Kameke2011} A. von Kameke, F. Huhn, G. Fern\'{a}ndez-Garc\'{i}a, A.P. Mu\~{n}uzuri, and V. P\'{e}rez-Mu\~{n}uzuri, \textit{Double Cascade Turbulence and Richardson Dispersion in a Horizontal Fluid Flow Induced by Faraday Waves}, Phys. Rev. Lett. \textbf{107}, 074502 (2011).
\bibitem{Francois2013} N. Francois, H. Xia, H. Punzmann, and M. Shats, \textit{Inverse Energy Cascade and Emergence of Large Coherent Vortices in Turbulence Driven by Faraday Waves}, Phys. Rev. Lett. \textbf{110}, 194501 (2013).
\bibitem{Gollub1990} R.Ramshankar, D.Berlin and J.P.Gollub, \textit{Transport by Capillary Waves. Part 1. Particle Trajectories}, Physics of Fluids A: Fluid Dynamics \textbf{2}, 1955 (1990).
\bibitem{Mesquita1992} O.N.Mesquita, S.Kane and J.P.Gollub, \textit{Transport by Capillary Waves: Fluctuating Stokes Drift}, Phys. Rev. A \textbf{45}, 3700 (1992).
\bibitem{Hansen1997} A.E. Hansen, E. Schr\"{o}der, P. Alstr{\o}m, J.S. Andersen, and M.T. Levinsen, \textit{Fractal Particle Trajectories in Capillary Waves: Imprint of Wavelength}, Phys. Rev. Lett. \textbf{79}, 1845 (1997).
\bibitem{Stokes1847} G.Stokes, \textit{On the Theory of Oscillatory Waves}, Trans Cambridge Philos Soc \textbf{8}, 441 (1847).
\bibitem{Hasselmann1982} K.Herterich and K. Hasselmann, \textit{The Horizontal Diffusion of Tracers by Surface Waves}, J.Phys. Oceanog. \textbf{12}, 704 (1982).
\bibitem{XiaNC2013} H. Xia, N. Francois, H. Punzmann, and M. Shats, \textit{Lagrangian Scale of Particle Dispersion in Turbulence}, Nat. Commun. \textbf{4} (2013).
\bibitem{Xia2014} H. Xia, N. Francois, H. Punzmann, and M. Shats, \textit{Taylor Particle Dispersion during Transition to Fully Developed Two-Dimensional Turbulence}, Phys. Rev. Lett. \textbf{112}, 104501 (2014).
\bibitem{Shats2012} M. Shats, H. Xia, and H. Punzmann, \textit{Parametrically Excited Water Surface Ripples as Ensembles of Oscillons}, Phys. Rev. Lett. \textbf{108}, 034502 (2012).
\bibitem{Taylor1921} G.I. Taylor, \textit{Diffusion by Continuous Movements}, Proc. London Math. Soc. \textbf{20}, 196--211 (1921).
\bibitem{Xia2012} H. Xia, T. Maimbourg, H. Punzmann, and M. Shats, \textit{Oscillon Dynamics and Rogue Wave Generation in Faraday Surface Ripples}, Phys. Rev. Lett. \textbf{109}, 114502 (2012).
\bibitem{Kraichnan1967} R.Kraichnan, \textit{Inertial Ranges in Two-Dimensional Turbulence}, Phys. Fluids \textbf{10}, 1417 (1967).
\bibitem{Bandi2006} M.M.Bandi, W.I.Goldburg, and J.R.Cressman Jr, \textit{Measurement of Entropy Production Rate in Compressible Turbulence}, EPL \textbf{76}, 595 (2006).
\bibitem{Larkin2009} J.Larkin, M.M.Bandi, A.Pumir, and W.I.Goldburg, \textit{Power-Law Distributions of Particle Concentration in Free-surface Flows}, PRE \textbf{80}, 066301 (2009).
\bibitem{Saffman} P.G.Saffman, \textit{Vortex Dynamics} (Cambridge Univ. Press, 1992).
\bibitem{L-Higgins1953} M.S.Longuet-Higgins, \textit{Mass Transport in Water Waves}, Phil. Trans. R. Soc. Lond. A \textbf{245}, 535 (1953).
\bibitem{L-Higgins1960} M.S.Longuet-Higgins, \textit{Mass Transport in the Boundary Layer at a Free Oscillating Surface}, J.Fluid Mech. \textbf{8}, 293 (1960).
\bibitem{Shrira1996} V.I.Shrira, and V.V.Voronovich, \textit{Nonlinear Dynamics of Vorticity Waves in the Coastal Zone}, Journal of Fluid Mechanics \textbf{326}, 181 (1996).
\bibitem{Crocker1996} J.C. Crocker, and D.G. Grier, \textit{Methods of Digital Video Microscopy for Colloidal Studies}, J. Colloid Interface Sci. \textbf{179}, 298--310 (1996).
\bibitem{Shats2005}  M.G. Shats, H. Xia, and H. Punzmann, \textit{Spectral Condensation of Turbulence in Plasmas and Fluids and its Role in Low-to-High Phase Transitions in Toroidal Plasma}, Phys. Rev. E \textbf{71}, 046409 (2005).


\end{thebibliography}
\end{document}